\documentclass[final,5p,times,fleqn]{elsarticle}
\usepackage{graphicx}
\usepackage{amssymb}
\usepackage{mathrsfs}
\usepackage{amsmath}
\usepackage{comment}
\usepackage{amsbsy}
\usepackage{mathrsfs}
\usepackage{amsfonts}
\usepackage{lineno}%(enumera las lineas)
\usepackage{nicefrac}
\usepackage{epsf,color,colordvi,pifont}
\usepackage[notref,notcite]{showkeys} %% just for draft versions

\DeclareFontFamily{OT1}{pzc}{}
\DeclareFontShape{OT1}{pzc}{m}{it}{<-> s * [1.10] pzcmi7t}{}
\DeclareMathAlphabet{\mathpzc}{OT1}{pzc}{m}{it}

\begin{document}

\begin{frontmatter}
\title{Photo-production  of scalar  particles  in the field of a  circularly  polarized laser beam}
\author{S. Villalba-Ch\'avez}
\ead{selym@tp1.uni-duesseldorf.de}

\author{C. M\"{u}ller}
\ead{c.mueller@tp1.uni-duesseldorf.de}
\address{Institut f\"{u}r Theoretische Physik I, Heinrich-Heine-Universit\"{a}t D\"{u}sseldorf, Universit\"{a}tsstr. 1, 40225 D\"{u}sseldorf, Germany}
\date{\today}

\begin{abstract}
The photo-production of a pair of scalar particles in the  presence of an intense, circularly
polarized laser beam  is investigated.  Using the optical theorem  within the framework of scalar 
quantum electrodynamics,  explicit expressions are given  for the pair production probability in terms of the  imaginary part of the vacuum polarization tensor. Its
leading asymptotic behavior is determined for various  limits of interest. The influence of the absence of 
internal spin degrees of freedom is analyzed via a comparison with the corresponding probabilities for 
production  of  spin-$\nicefrac{1}{2}$ particles; the lack of spin is shown to suppress the pair creation 
rate, as compared to the predictions from Dirac theory. 
Potential applications of our results for the search of minicharged particles are indicated.
% Possible experimental scenarios where  our predictions could be tested   are also discussed.
\end{abstract}
\begin{keyword}
Vacuum Polarization \sep Laser Fields \sep Particle Production \sep Spin Effects.
\PACS 12.20.Ds \sep 11.10.Jj \sep 13.40-f \sep  32.80.Wr.
\end{keyword}
\end{frontmatter}

%%%%%%%%%%%%%%%%%%%%%%%%%%%%%%%%%%%%%%%%%%%%%%%%%%%%%%%%%%%%%%%%%%%%%%%%%%%%%%%%%%%%%%%%%%%%%%%%%%%%%%%%%%%%%%%%%%%%%%%%%%%%%%%%%%%%%%
\section{Introduction}
%%%%%%%%%%%%%%%%%%%%%%%%%%%%%%%%%%%%%%%%%%%%%%%%%%%%%%%%%%%%%%%%%%%%%%%%%%%%%%%%%%%%%%%%%%%%%%%%%%%%%%%%%%%%%%%%%%%%%%%%%%%%%%%%%%%%

Understanding the nonlinear and unstable nature of the quantum vacuum in the presence of a strong  electromagnetic field  constitutes an important
task of  theoretical physics. Corresponding studies have revealed  a nontrivial vacuum structure,  suitable  to explore the low-energy frontier  of
particle physics \cite{Gies:2008wv,Jaeckel:2010ni,Dobrich:2012sw}. Moreover,  perspectives of achieving ultrahigh field  intensities
($I \sim 10^{26}$  $\mathrm{W/cm}^2$) in  short laser pulses of few femtoseconds duration \cite{ELI,hiper} have  motivated  a growing interest in 
the  phenomenology  purely  associated with the quantum nature of the electromagnetic interaction (see \cite{Ehlotzky:2009,DiPiazza:2011tq} for recent 
reviews). This is because  the envisaged laser field strengths lie only 1-2 orders of magnitude below the critical value $E_{\mathrm{c}}=1.3\times 10^{16}\ \mathrm{V/cm}$  
where QED vacuum nonlinearities become substantial and spontaneous vacuum decay into electron-positron ($e^-e^+$) pairs via the Schwinger mechanism 
is expected to occur \cite{Sauter:1931,Heisenberg:1935,Schwinger:1951nm}.

In combination with an incident high-energy particle, strong laser fields can induce $e^-e^+$ pair production already at intensities available today. 
In a pioneering experiment at SLAC \cite{Burke:1997ew}, a multi-GeV photon decayed into a pair while propagating through a moderately intense laser 
pulse ($I\sim 10^{18}$  $\mathrm{W/cm}^2$). This process, involving the simultaneous absorption of several laser photons, represents a nonlinear 
version of the well-known Breit-Wheeler reaction \cite{Breit:1934zz,Reiss1962,narozhnyi}. The high-energy non-laser photon originated from Compton 
backscattering of SLAC's ultrarelativistic electron beam off the laser pulse. In the near future, corresponding studies can be conducted within all-optical 
setups using laser-accelerated relativistic electrons as projectiles \cite{huayuhu}. Other pair production mechanisms may be probed in ultrarelativistic 
proton-laser collisions \cite{Muller:2003zz,Sieczka:2006,Milstein:2006zz,DiPiazza:2009py,Muller:2008ys}.

In view of the upcoming high-field laboratories \cite{ELI,hiper}, theoreticians are currently investigating further properties and applications of photo-induced 
$e^-e^+$ pair production in intense laser fields. For example,  due to their broad frequency composition, laser pulses of ultrashort duration have been 
shown to modify the created particle spectra \cite{Heinzl:2010vg} and lead to characteristic enhancements in the pair production probability \cite{Titov,Nousch}. 
Photo-induced pair production also plays a crucial role for the development of QED cascades which may give rise to $e^-e^+$ plasmas of very high density 
\cite{Bell:2008,Nerush:2011}. The nonlinear Breit-Wheeler process moreover offers a promising means to measure ultrashort $\gamma$-ray pulses via $e^-e^+$ 
streaking \cite{Ipp}. Superimposing the field of a high-energy photon onto a strong electric field may also help catalyzing the Schwinger effect \cite{Schutzhold:2008pz,Dunne:2009gi,Orthaber:2011cm}.

In the present Letter, we study the photo-induced creation of a pair of spin$-0$ particles in the presence of a strong monochromatic laser beam. Our 
motivation is twofold. First, while the probabilities for the creation of fermion pairs are known for a long time \cite{Reiss1962,narozhnyi}, it is 
relevant to establish the corresponding formulas for scalar particles because they can be useful for the ongoing search  of minicharged  particles 
which may have either fermionic or bosonic character \cite{Jaeckel:2010ni,Dobrich:2012sw,Ahlers:2006iz}. Second, our results provide insights into the 
fundamental question as to how the spin degree of freedom affects the photo-induced pair production process. To this end, a comparison with the known 
results for fermion pair production will be drawn. Such an information complements previous works where spin-resolved calculations of the nonlinear 
Breit-Wheeler process via helicity amplitudes \cite{Tsai:1992ek} and the internal spin polarization vector \cite{Ivanov:2004vh} were performed. We note 
besides that comparative studies between the behavior of bosonic and fermionic particles in strong laser fields have recently been carried out with 
respect to Compton, Mott and Kapitza-Dirac scattering \cite{Ehlotzky:2009,Boca:2012,Ahrens:2012}, nonlinear Bethe-Heitler pair creation in proton-laser 
collisions \cite{Muller:2010xq} and the Klein paradox \cite{Grobe_Spin2010,Krekora:2004}. 

Our theoretical approach relies on the polarization tensor $\Pi_{\mu\nu}(k_1,k_2),$ of scalar 
Quantum Electrodynamics  (in the one-loop approximation)  in the presence of a strong laser field 
\cite{baier,baierbook}  whose imaginary part is related to the pair production probability  of scalar particles  via the optical theorem. While the polarization tensor 
for Dirac fermions has already been exploited successfully to calculate various $e^-e^+$ pair production processes in strong laser fields 
\cite{Milstein:2006zz,DiPiazza:2009py,baier}, to the best of our knowledge the present calculations represent the first application of the corresponding 
polarization tensor for the scalar case.

%%%%%%%%%%%%%%%%%%%%%%%%%%%%%%%%%%%%%%%%%%%%%%%%%%%%%%%%%%%%%%%%%%%%%%%%%%%%%%%%%%%%%%%%%%%
\section{General considerations \label{GC}}
%%%%%%%%%%%%%%%%%%%%%%%%%%%%%%%%%%%%%%%%%%%%%%%%%%%%%%%%%%%%%%%%%%%%%%%%%%%%%%%%%%%%%%%%%%%

To begin with, let us consider the field of a plane electromagnetic wave of the form\footnote{From now on ``natural'' and Gaussian  units  $c=\hbar=4\pi\epsilon_0=1$ are used.}
\begin{equation}
\begin{split}
 &\mathscr{A}^{\mu}(x)=\mathpzc{a}_1^{\mu}\psi_1(\varkappa x)+\mathpzc{a}_2^{\mu}\psi_2(\varkappa x),
 \end{split}
\end{equation} 
with $\mathpzc{a}_{1,2}$ denoting the wave amplitudes and $\psi_{1,2}$ being arbitrary functions. The wave four-vector $\varkappa^{\mu}=(\varkappa^0, \pmb{\varkappa})$ 
fulfills the relations $\varkappa^2=0$ and $\varkappa \mathpzc{a}_1=\varkappa \mathpzc{a}_2=\mathpzc{a}_1\mathpzc{a}_2=0$. According to \cite{baier,baierbook}, the vacuum 
polarization tensor in this field, 
\begin{eqnarray}\label{DVPBaier}
\Pi^{\mu\nu}(k_1,k_2)&=&c_1\Lambda^\mu_{1}\Lambda^\nu_{2}+c_2\Lambda^\mu_{2}\Lambda^\nu_{1}+c_3\Lambda^\mu_{1}\Lambda^\nu_{1}\nonumber\\&&+c_4\Lambda^\mu_{2}\Lambda^\nu_{2}+c_5\Lambda^\mu_{3}\Lambda^\nu_{4}
\end{eqnarray}
can be expanded in terms of a basis set of Lorentz covariant vectors $\Lambda^\mu_{i}$ which are constructed from fundamental symmetry principles. They are explicitly given by
\begin{eqnarray}\label{vectorialbasisbaier}
\begin{array}{c}\displaystyle
\Lambda_{1}^\mu(k)=-\frac{\mathscr{F}_{1}^{\mu\nu}k_\nu}{(k\varkappa)\left(-\mathpzc{a}_1^2\right)^{\nicefrac{1}{2}}},\quad  \Lambda_{2}^\mu(k)=-\frac{\mathscr{F}_{2}^{\mu\nu}k_\nu}{(k\varkappa)\left(-\mathpzc{a}_2^2\right)^{\nicefrac{1}{2}}},\\ \\ \displaystyle
\Lambda_{3}^\mu(k)=\frac{\varkappa^\mu k_1^2-k_{1}^{\mu} (k\varkappa)}{(k\varkappa)\left(k_1^2\right)^{\nicefrac{1}{2}}},\quad  \Lambda_{4}^\mu(k)=\frac{\varkappa^\mu k_2^2-k_{2}^{\mu} (k\varkappa)}{(k\varkappa)\left(k_2^2\right)^{\nicefrac{1}{2}}}.
\end{array}
\end{eqnarray}Here $\mathscr{F}^{\mu\nu}_{i}=\varkappa^\mu\mathpzc{a}^\nu_{i}-\varkappa^\nu\mathpzc{a}^\mu_{i}$ $(i=1,2)$ are the amplitudes of the  external 
field modes whereas $k_1$ and $k_2$ denote  the  incoming  and outgoing four-momenta of the probe photons, respectively. We note that the short-hand notation $k$ in Eq.~\eqref{vectorialbasisbaier} may stand for either $k_1$ or $k_2.$  It is worth mentioning at this point 
that, for $k=k_1$, the  vectors $\Lambda_1(k_1),$ $\Lambda_2(k_1)$ and $\Lambda_3(k_1)$ are orthogonal to each other,  $\Lambda^\mu_{i}(k_1)\Lambda_{j\mu}(k_1)=-\delta_{ij}$,
and fulfill the completeness relation $
g^{\mu\nu}-\frac{k_1^{\mu} k_1^{\nu}}{k_1^2}=-\sum_{i=1}^{3}\Lambda^\mu_{i}(k_1)\Lambda^\nu_{i}(k_1)$ with $g_{\mu\nu}=\mathrm{diag}(+1,-1,-1,-1)$ 
denoting the metric tensor. A similar statement applies if the set of vectors   $\Lambda_1(k_2),$ $\Lambda_2(k_2)$ and $\Lambda_4(k_2)$ are considered.
We emphasize that Eq.~\eqref{DVPBaier} does not depend on which choice of $k$ is taken since the difference between $k_1$ and $k_2$ is proportional 
to $\varkappa$ [see Eq.~\eqref{pimunueigenavluesva} below].

The form factors  $c_i$ in  Eq. (\ref{DVPBaier}) are distribution-valued functions which depend on the field shape via the functions 
$\psi_i$. They have been evaluated thoroughly for the case of spin$-\frac{1}{2}$ particles in \cite{baier,baierbook}. Also for the case 
when the virtual charge carriers in the Feynman loop are spin-$0$ particles general expressions for the $c_i$ were provided in these references; but these 
formulas were not further evaluated.

Using the general expressions from \cite{baier,baierbook} and assuming that the laser field is an elliptically polarized wave with
\begin{equation}
\psi_1=\cos(\varkappa x)\quad \mathrm{and} \quad \psi_2=\sin(\varkappa x),
\end{equation}we find that the form factors in  Eq. (\ref{DVPBaier}) for the scalar case are given by
\begin{eqnarray}
&&c_i=-i\frac{\alpha}{\pi}m^2\int_{-1}^1dv\int_0^\infty \frac{d\rho}{\rho}e^{-\frac{2i\rho}{\vert\lambda\vert\left(1-v^2\right)}\left(1+A\left(\xi_1^2+\xi_2^2\right)-\frac{k_1^2\left(1-v^2\right)}{4m^2}\right)}\nonumber\\ &&\quad\times(2\pi)^4\left[\delta^4(k_1-k_2)d_i^{(0)}+\sum_{\stackrel{N=-\infty}{N\neq0}}^\infty\delta^4(k_1-k_2-2N\varkappa)d_i^{(N)}\right].
\label{pimunueigenavluesva}
\end{eqnarray}Here, $\alpha=e^2$ is the fine structure constant, $e$ and $m$ denote the particle charge and mass, respectively, and  
\begin{equation}
\lambda=\frac{\varkappa k}{2m^2},\quad \xi_i^2=-\frac{e^2\mathpzc{a}_i^2 }{m^2}\quad \ (i=1,2).
\label{xi}
\end{equation}
As Eq.~\eqref{pimunueigenavluesva} shows, the polarization tensor decomposes into elastic ($k_1=k_2$) and inelastic ($k_1\ne k_2$) parts. Those  terms which  
contain the  Dirac deltas  $\delta^4(k_1-k_2+2N\varkappa)$ with $N\neq0$  are responsible for the  inelastic scattering of a photon in the field of the wave. 
For our purposes, however, only the elastic part is relevant. The corresponding functions $d^{(0)}_i,\ i=1,2,3,4,5$  contained in Eq.~(\ref{pimunueigenavluesva}) are given by
\begin{eqnarray}
& d_1^{(0)}=-d_2^{(0)}= \xi_1\xi_2 \rho A_0 J_0(z)\mathrm{sign}[\lambda], \\
& d_3^{(0)}=-\frac{1}{2} \xi_1^2 A_1\left(J_0(z)-iJ_0^\prime(z)\right)+\frac{\xi_1^2}{2}\sin^2(\rho)J_0(z)\nonumber\\
&\qquad\quad+\frac{i\vert\lambda\vert\left(1-v^2\right)}{8\rho}\left(J_0(z)-e^{iy}\right),\\
& d_4^{(0)}=d_3^{(0)}\left(\xi_1^2\leftrightarrow\xi_2^2\right),\  d_5^{(0)}=-\frac{k_1^2}{8m^2}v^2 \left(J_0(z)-e^{iy}\right)
\end{eqnarray}where $J_0(z)$ is the Bessel function of zero order and $J_0^\prime(z)$ its derivative. The remaining  parameters  are
\begin{eqnarray}
\begin{array}{c}\displaystyle
A=\frac{1}{2}\left(1-\frac{\sin^2(\rho)}{\rho^2}\right),\quad A_0=\frac{1}{2}\left(\frac{\sin^2(\rho)}{\rho^2}-\frac{\sin(2\rho)}{2\rho}\right), \\ \\  \displaystyle
z=\frac{2\rho A_0}{1-v^2}\frac{\xi_1^2-\xi_2^2}{\vert\lambda\vert},\quad  \displaystyle y=\frac{2\rho A}{1-v^2}\frac{\xi_1^2+\xi_2^2}{\vert\lambda\vert},
\end{array}
\label{parameters}
\end{eqnarray}and  $A_1=A+2A_0.$  

A substantial simplification is achieved when the external field is taken as a circularly polarized wave ($\xi_1=\xi_2=\xi$). In this case,  we find it 
convenient to express  the elastic contribution as
\begin{equation}
\Pi_{\mu\nu}^{(\mathrm{elast})}(k_1,k_2)=i(2\pi)^4\delta^4(k_1-k_2)\Pi_{\mu\nu}(k_2)
\end{equation}with
\begin{eqnarray}
&&\Pi^{\mu\nu}(k_2)=\left(\Lambda^\mu_{1}\Lambda^\nu_{2}-\Lambda^\mu_{2}\Lambda^\nu_{1}\right)\pi_1^{(0)}+
\left(\Lambda^\mu_{1}\Lambda^\nu_{1}+\Lambda^\mu_{2}\Lambda^\nu_{2}\right)\pi_3^{(0)}\nonumber\\ &&\qquad\qquad+\Lambda^\mu_{3}\Lambda^\nu_{3}\pi_5^{(0)}.
\end{eqnarray}Here the involved coefficients are given by
\begin{eqnarray}\label{circulareigenavlues1}
\pi_i^{(0)}=-\frac{\alpha}{2\pi}m^2\int_{-1}^1dv\int_0^\infty \frac{d\rho}{\rho}e^{-\frac{2i\rho}{\vert\lambda\vert\left(1-v^2\right)}\left(1\frac{}{}+2A\xi^2
-\frac{k_2^2\left(1-v^2\right)}{4m^2}\right)}\Omega_i^{(0)}
\end{eqnarray}where
\begin{eqnarray}
&\Omega_1^{(0)}=2\xi^2 A_0\rho\mathrm{sign}\left[\lambda\right],\ \ \Omega_5^{(0)}=-\frac{k_2^2}{4m^2}v^2\left(1-e^{iy}\right), \\
&\Omega_3^{(0)}=\xi^2\sin^2\left(\rho\right)+\frac{1}{2}\left[1-\frac{k_1 k_2}{4m^2}(1-v^2)\right]\left(1-e^{iy}\right).
\label{circulareigenavlues2}
\end{eqnarray}

\begin{figure}
\includegraphics[width=3.5in]{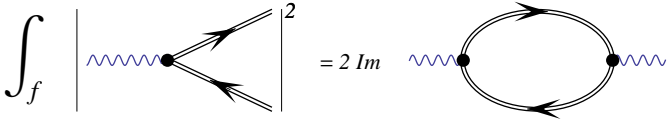}
\caption{\label{fig:mb000} Diagrammatic representation of the  optical theorem  applied to the photo-production of a pair of spinless particles  in the field of a wave. 
In the left-hand side we have  represented  the squared modulus  of the the $S-$matrix element  integrated  over the final phase volume 
$\int_f\equiv \int \frac{d^3p_{+}}{(2\pi)^3}\frac{d^3p_{-}}{(2\pi)^3}$.  Here the double lines refer to the  exact  Klein-Gordon states interacting with the external background. 
The right-hand side, however,  contains the imaginary part of the vacuum polarization tensor where the  double lines  represent the exact  propagators in the field of the wave. 
In both sides the  wavy lines denote the legs corresponding to a   photon field.}
\end{figure}

Now, the unitarity condition of the dispersion $S-$matrix provides the optical theorem. According to the
latter, the  total creation rate  of  a  pair of spin-$0$ particles  from a real photon $(k^2=0)$ with
polarization vector $\epsilon_\ell^\mu$ $(\ell=1,2)$ turns out to be  $\mathfrak{R}_{\ell}^{(0)}=\epsilon^{\mu}_\ell\epsilon^{*\nu}_{\ell}\rm Im\ \Pi_{\mu\nu}/\omega$   where $\omega$ 
is the photon frequency (see Fig. \ref{fig:mb000}). If the photon is unpolarized, the averaged rate $\mathfrak{R}^{(0)}=\left(\mathfrak{R}_1^{(0)}+\mathfrak{R}_2^{(0)}\right)/2$ reads 
\begin{equation}\label{paircreapro}
\mathfrak{R}^{(0)}=\sum_{\ell}\frac{\epsilon_\ell^\mu\epsilon_\ell^{*\nu}}{2\omega}\mathrm{Im}\ \Pi_{\mu\nu}(k)=-\frac{g^{\mu\nu}}{2\omega}\mathrm{Im}\ \Pi_{\mu\nu}(k)\,,
\end{equation} where the completeness relation $g^{\mu\nu}=-\sum_{\ell}\epsilon_\ell^\mu\epsilon^{*\nu}_{\ell}$ was used in the second step. 
Because of this fact, the expression above reduces to
\begin{equation}
\mathfrak{R}^{(0)}=\frac{\mathrm{Im}\ \pi_3^{(0)}}{\omega}.
\end{equation}So, among the set of form factors previously described in Eqs. (\ref{circulareigenavlues1})-(\ref{circulareigenavlues2}),
only one contributes to $\mathfrak{R}^{(0)}.$ In correspondence, the probability of producing a pair of scalar particles by a photon traveling 
through a laser field of circular polarization can be represented by the following double-parametric integral
\begin{eqnarray}\label{scalarproba}
&&\mathfrak{R}^{\left(0\right)}(\xi,\lambda)=-\frac{\alpha m^2}{2\pi\omega}\mathrm{Im}\ \int_0^1dv\int_0^\infty \frac{d\rho}{\rho}e^{-\frac{2i\eta}{1-v^2}}
\left\{1-e^{iy}\right. \nonumber\\ &&\qquad\qquad\qquad+\left.2\xi^2\sin^2(\rho)\right\},
\end{eqnarray}where the abbreviation  $\eta=(\rho/\lambda)\left[1+\xi^2\left(1-\sin^2(\rho)/\rho^2\right)\right]$ has been
used.  In the case under consideration  the  parameter $y$ of Eq. (\ref{parameters}) becomes $y=\frac{2\xi^2}{1-v^2} \frac{\rho}{\lambda}\left(1-\sin^2(\rho)/\rho^2\right).$  
Note that for a real non-laser photon (with  $k^2=0$,  i.e.  $\omega=\vert\pmb{k}\vert$) the parameter $\lambda$ in Eq.~\eqref{xi} is always nonnegative $\lambda\geqslant0$.

The structure of Eq. (\ref{scalarproba}) shares certain similarities with the corresponding production rate $\mathfrak{R}^{(\frac{1}{2})}$ of 
spin-$\frac{1}{2}$ particles which is given by \cite{baierbook}:
\begin{eqnarray}\label{spinorproba}
&&\mathfrak{R}^{\left(\frac{1}{2}\right)}(\xi,\lambda)=\frac{\alpha m^2}{\pi\omega}\mathrm{Im}\ \int_0^1dv\int_0^\infty \frac{d\rho}{\rho}e^{-\frac{2i\eta}{1-v^2}}
\left\{1-e^{iy}\right. \nonumber\\ &&\qquad\qquad\qquad-\left.2\xi^2\frac{1+v^2}{1-v^2}\sin^2(\rho)\right\}.
\end{eqnarray}Eqs. (\ref{scalarproba}) and (\ref{spinorproba})  differ, however, by (i) an overall  factor of $-2$ which coincides with the 
spin$-\frac{1}{2}$ multiplicity, and (ii) the precise form of the integral over $v.$  As we will see, this gives rise to differences between $\mathfrak{R}^{(0)}$
and $\mathfrak{R}^{(\frac{1}{2})}$ regarding both the absolute size and the functional dependence. We should mention at this point that an exact 
evaluation of  $\mathfrak{R}^{(0)}$ is quite difficult to perform.  However, the dependence on the parameters of the theory,  i.e.  $\lambda$ 
and $\xi$, allows to obtain closed-form analytical rate expressions in various asymptotic regimes of interest. They will be derived in the 
forthcoming section.

%%%%%%%%%%%%%%%%%%%%%%%%%%%%%%%%%%%%%%%%%%%%%%%%%%%%%%%%%%%%%%%%%%%%%%%%%%%%%%%%%%%%%%%%%%%%%%%%%%%%%%%%
\section{Asymptotic regimes and comparison with spin-$\frac{1}{2}$ case \label{Asymp}}
%%%%%%%%%%%%%%%%%%%%%%%%%%%%%%%%%%%%%%%%%%%%%%%%%%%%%%%%%%%%%%%%%%%%%%%%%%%%%%%%%%%%%%%%%%%%%%%%%%%%%%%%

Before undertaking the main calculations of this section,  we briefly  analyze  the kinematics associated 
with  the decay process $\gamma(k) + n\gamma_L(\varkappa)\to \ell^-+\ell^+$, where $n$ denotes the number of
participating laser photons $\gamma_L$.  In the center-of-mass frame, the corresponding energy-momentum balance implies 
the condition $n k\varkappa=2\varepsilon^2$, with $\varepsilon$ being the (laser-dressed) energy of a final particle 
state. Accordingly,  whenever  the  number of absorbed laser photons  exceeds  the threshold  value
\begin{equation}
n_*=\frac{2m_*^2}{k\varkappa}\label{threshold}
\end{equation} the decay can occur. Here,  $m_*\equiv m (1+\xi^2)^{\nicefrac{1}{2}}$ denotes the effective 
mass of the particle. It comes out as a consequence  of considering the dressed four-momentum 
$q_\mu=p_\mu+\frac{m^2\xi^2}{(\varkappa p)}\varkappa_\mu$  
(with $p_{\mu}$ the free four-momen\-tum of the particle, i.e. $p^2=m^2$)  as the kinematical variable 
involved in the energy-momentum conservation \cite{Reiss1962,narozhnyi}. We point out that in the 
center-of-mass frame, the relative speed between the created particles  is given by  
$\vert\pmb{\mathpzc{v}}_{\mathrm{rel}}\vert=\vert\pmb{\mathpzc{v}}_{-}-\pmb{\mathpzc{v}}_{+}\vert=2\mathpzc{v}$   with
\begin{equation}
\mathpzc{v}=\frac{\vert\pmb{q}\vert}{\varepsilon}=\left(1-\frac{n_*}{n}\right)^{\nicefrac{1}{2}}. 
\end{equation}

%%%%%%%%%%%%%%%%%%%%%%%%%%%%%%%%%%%%%%%%%%%%%%%%%%%%%%%%%%%%%%%%%%%%%%%%%%%%%%%%%%%%%%%%%%%%%%%%%%%%%%%%
\subsection{Two-photon reaction at $\xi< 1$}
%%%%%%%%%%%%%%%%%%%%%%%%%%%%%%%%%%%%%%%%%%%%%%%%%%%%%%%%%%%%%%%%%%%%%%%%%%%%%%%%%%%%%%%%%%%%%%%%%%%%%%%%

We wish to specialize Eq. (\ref{scalarproba}) to the case where $\xi < 1.$
In this context, we note that the  probability  of creating  a pair is  suppressed if the  condition
$\lambda\ll1$ holds, so that dispersive processes become of main interest here\footnote{In the aforementioned
limit, the theoretical description of such phenomena can be performed by means of an  effective Lagrangian approach. 
For details we refer the reader to \cite{DiPiazza:2011tq,Heinzl,BenKing:2009,VillalbaChavez:2012ea}.}.
On the contrary, if   $\lambda\geqslant1+\xi^2$  (i.e. $n_*\leqslant1$), the pair production could take place  with the
absorption of just one photon from the laser wave (two-photon reaction).  In this case, which we shall consider now, 
the   oscillatory contribution present
in  the exponent of  (\ref{scalarproba}) becomes very small in comparison with the  remaining terms,
allowing to expand  $\exp\left[\frac{i\xi^2}{\lambda} \frac{\sin^2(\rho)}{\rho}\right]$ without
obstruction. Afterwards, we perform a  change of variable %according to 
$\rho\to \rho(1+\xi^2)$ in the integral which corresponds to the  second term inside the curly brackets in Eq. (\ref{scalarproba}) to  arrive at
%in the integral $\int_0^{\infty}d\rho e^{- \frac{2i\rho}{\lambda\left(1-v^2\right)}}/\rho$ that we end up with
\begin{eqnarray}
&&\mathfrak{R}^{(0)}\backsimeq-\frac{\alpha m^2\xi^2}{2\pi\omega}\ \int_0^1dv\int_{-\infty}^\infty \frac{d\rho}{\rho}e^{\frac{2i\rho \left(1+\xi^2\right)}{\lambda\left(1-v^2\right)}}\sin^2(\rho)\nonumber\\ &&\qquad\qquad\times
\left\{i+\frac{1}{\lambda\rho\left(1-v^2\right)}\right\}\label{scalarprobaborn0}.
\end{eqnarray} In the derivation of  this expression a term  of the order $\sim\xi^2/\lambda$  has been accounted for, 
whereas terms of higher order in $\xi^2/\lambda$ were dropped.  Hence, the results obtained   in this section  apply  whenever 
the following conditions  are simultaneously fulfilled:
\[
\lambda\geqslant1+\xi^2\qquad\mathrm{and}\qquad \lambda\gg\xi^2.
\]Besides, to obtain Eq . (\ref{scalarprobaborn0})  the $\mathrm{Im} (\ldots)$  present in Eq. (\ref{scalarproba}) must  be  carried   out,  after which   
the resulting integrand  turns  out to be an even function in the $\rho-$variable.  This symmetry allows  to 
perform  the change in the integration limits according to   $\int_0^\infty d\rho\ldots\to\frac{1}{2}\int_{-\infty}^\infty d\rho\ldots$ 
as well as to express the integrand  as it stands in (\ref{scalarprobaborn0}).

In order to provide  an explicit expression of $\mathfrak{R}^{(0)}$  we integrate by parts the terms
containing the  factor proportional to $1/\rho^2.$ Afterwards, the residue theorem  is   applied.  
To this end,  the  contour  of the $\rho-$integration is  chosen slightly below the real $\rho$ axis 
(cf.  also  \cite{baierbook}). As a consequence we obtain
\begin{eqnarray}
&&\mathfrak{R}^{(0)}\backsimeq-\frac{\alpha m^2\xi^2}{2\omega}\ \int_0^1dv\ \theta\left(1-\frac{1+\xi^2}{\lambda\left(1-v^2\right)}\right)
\nonumber\\ &&\qquad\quad\times
\left\{\frac{1}{\lambda\left(1-v^2\right)}-\frac{1}{2}\left(1+\frac{2\left(1+\xi^2\right)}{\lambda^2\left(1-v^2\right)^2}\right)\right\}\label{scalarprobaborn}
\end{eqnarray}where   $\theta(x)$ denotes the unit step function.  The remaining  integral over $v$ can be
taken analytically without  complications. It leads to
\begin{eqnarray}
&&\mathfrak{R}^{(0)}\backsimeq\frac{\alpha m^2\xi^2}{4\omega}\left[\left(1+\frac{n_*}{1+\xi^2}\right)\sqrt{1-n_*}-\frac{n_*}{1+\xi^2}\left(1-\frac{1}{2}n_* \right)\right.\nonumber\\
&&\qquad\quad\qquad\times\left.\ln\left(\frac{1+\sqrt{1-n_*}}{1-\sqrt{1-n_*}}\right)\right]\theta(1-n_*)\label{scalarprobabornsolu}
\end{eqnarray}with $n_*$ given in   (\ref{threshold}). The respective expression for $\mathfrak{R}^{\left(\frac{1}{2}\right)}$ 
can be read off from Eq. (\ref{scalarprobabornsolu}) by multiplying the latter by $-2$ and replacing the coefficient in front of 
the logarithmic function by $1+n_*/(1+\xi^2)-n_*^2/2(1+\xi^2)$ (see Appendix D in \cite{baierbook}). 

Let us consider some limiting cases. At $n_*\backsimeq 1$, i.e. in the nonrelativistic limit ($\mathpzc{v}\ll1$), Eq. (\ref{scalarprobabornsolu}) behaves as
\begin{equation}\label{nonrelativistic}
\mathfrak{R}^{(0)}\approx\frac{\alpha m^2\xi^2}{4\omega}\left(1-n_*\right)^{\nicefrac{1}{2}},
\end{equation} whereas $\mathfrak{R}^{\left(\frac{1}{2}\right)}= 2\mathfrak{R}^{\left(0\right)}.$  So, in this  limit,  the ratio between the  
production probabilities  coincides with the spin  multiplicity of a Dirac particle. Incidentally,   this relation is also manifest to leading 
order between  the rates of the respective  Schwinger mechanisms \cite{Dunne:2004nc}. In contrast, for  $n_*\approx 0,$  the  created particles are  
ultrarelativistic  $(\mathpzc{v}\sim1)$ and  the  probability  becomes   independent of  $\lambda:$
\begin{equation}
\mathfrak{R}^{(0)}\approx\frac{\alpha m^2\xi^2}{4\omega}. \label{lastfer}
\end{equation}The situation is quite different when the photo-production of  Dirac particles is considered. In fact, under  the same circumstances, 
the creation rate  of a pair of spin-$\frac{1}{2}$ particles has a   logarithmic dependence: 
$\mathfrak{R}^{\left(\frac{1}{2}\right)}\approx\frac{\alpha m^2\xi^2}{2\omega}\left[ 2\ln\left(2\varepsilon/m_*\right)-1\right];$ see also  
\cite{Greiner}. As a consequence,  the  ratio  between the fermionic and sca\-lar rates grows  logarithmically, i.e.  
$\mathfrak{R}^{\left(\frac{1}{2}\right)}/\mathfrak{R}^{(0)}\sim 4\ln\left(2\varepsilon/m_*\right)$  as $\varepsilon\sim(k\varkappa)^{1/2}\gg m_*.$ 
Such  a  behavior resembles the corresponding result arising  in Compton scattering. Indeed,  as is well known, when the energy of the incoming 
photon is very large $(\omega\gg m)$ the total Compton cross section  computed for Dirac fermions turns out to be   
$\sigma^{\left(\frac{1}{2}\right)}\approx \frac{\alpha^2\pi}{\omega m}\ \left[\ln\left(2\omega/m\right)+1/2\right]$. In contrast, the leading 
asymptotic behavior of the corresponding total cross section determined for  spin-$0$ particles reads $\sigma^{\left(0\right)}\approx \frac{2\alpha^2\pi}{\omega m}$ \cite{Greiner}. 
Accordingly, $\sigma^{\left(\frac{1}{2}\right)}/\sigma^{\left(0\right)}\sim \frac{1}{2}\ln\left(2\omega/m\right)$ as  $\omega\gg m$. 
Obviously, in the case under consideration,  $\mathfrak{R}^{\left(\frac{1}{2}\right)}/\mathfrak{R}^{(0)}$ can be substantially larger than the 
statistical factor of $4$  associated with  the possible   spin configurations in the final state.  

\begin{figure}
\includegraphics[width=3.in]{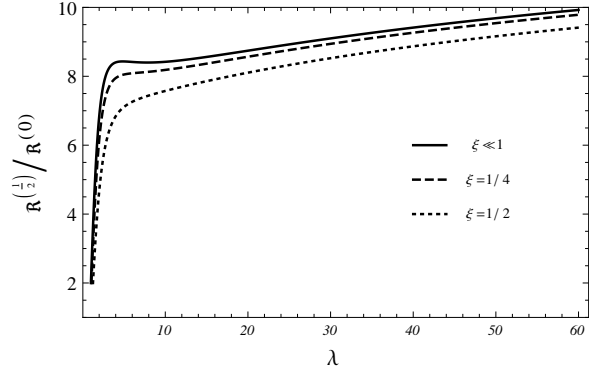}
\caption{\label{fig:mb001} Relative dependence  of  the fermionic and scalar pair production rates   on the parameter $\lambda$
for different values of $\xi<1$, as indicated. The case $\xi\ll1$ corresponds to the Born approximation.}
\end{figure}

The results obtained in this section    are summarized in  Fig.~\ref{fig:mb001} which   displays the ratio
$\mathfrak{R}^{\left(\frac{1}{2}\right)}/\mathfrak{R}^{\left(0\right)}$ as  a function of $\lambda.$ Several curves are shown corresponding to
different values of $\xi.$ The picture also includes  the limiting case of $\xi\ll1$ (solid  line), which is compatible with the Born 
approximation. The explicit expressions associated  with  this limit can be read off from Eqs. (\ref{scalarprobabornsolu})-(\ref{nonrelativistic}) by
replacing $n_*\to n_\mathrm{B}=1/\lambda$ and  setting $\xi=0$ within the squared brackets. The latter procedure leaves us with a   
quadratic dependence on the parameter $\xi.$ Moreover we note that, at larger values of $\xi\approx 1$ and $\lambda\sim 1+\xi^2,$ next-to-leading order terms become 
increasingly important (see also \cite{Milstein:2006zz}). These terms which   may  give some minor contribution 
to the rates for $\xi=1/2$ have not been included in Fig.~\ref{fig:mb001}. 

%%%%%%%%%%%%%%%%%%%%%%%%%%%%%%%%%%%%%%%%%%%%%%%%%%%%%%%%%%%%%%%%%%%%%%%%%%%%%%%%%%%%%%%%%%%%%%%%%%%%%%%%
\subsection{Leading behavior of $\mathfrak{R}^{(0)}$ at asymptotically large $\xi\gg1$}
%%%%%%%%%%%%%%%%%%%%%%%%%%%%%%%%%%%%%%%%%%%%%%%%%%%%%%%%%%%%%%%%%%%%%%%%%%%%%%%%%%%%%%%%%%%%%%%%%%%%%%

In order to undertake the  calculations in the high-field domain with $\xi\gg1$ it is convenient to  express  the integral of the first two terms in  Eq.~(\ref{scalarproba}) in a more suitable form:
\begin{eqnarray}
&&\mathrm{Im}\ \int_0^1dv\int_0^\infty\frac{d\rho}{\rho}e^{-\frac{2i\eta}{\left(1-v^2\right)}}(1-e^y)\nonumber\\
&&\ =-\frac{16\xi^2}{\lambda}\mathrm{Re}\ \int_0^1dv\frac{v^2}{\left(1-v^2\right)^2}\int_0^\infty d\rho A_0e^{-\frac{2i\eta}{\left(1-v^2\right)}}\nonumber\\
&&\ =-\frac{16\xi^2}{\lambda}\int_0^1\frac{dv v^2}{\left(1-v^2\right)^2}\int_{0}^\infty d\rho A_0\cos\left(\frac{2\eta}{\left(1-v^2\right)}\right)
\end{eqnarray}with  $A_0$  given   in  Eq. (\ref{parameters}) and $\eta$ defined below Eq. (\ref{scalarproba}). Inserting the expression above  into Eq. (\ref{scalarproba}) we obtain
\begin{eqnarray}\label{modified}
\mathfrak{R}^{(0)}=-\frac{i\alpha m^2}{4\pi\omega}\int_1^\infty\frac{du}{2u\left[u(u-1)\right]^{\nicefrac{1}{2}}}\int_{-\infty}^\infty \frac{d\rho}{\rho}e^{2i u\eta}\nonumber\\ \times\left\{2\xi^2\sin^2(\rho) +\frac{16 i\xi^2}{\lambda} u(u-1)\rho A_0\right\},
\end{eqnarray} where the lower boundary of the $\rho$-integral was extended to $-\infty$ taking into account the symmetry of the integrand  in this variable.
Besides, the change of  variable $u=\left(1-v^2\right)^{-1}$ has been carried out.
We  point out  that the integral over this variable  does not diverge, although the integrand is a singular function
at $u=1.$ Since the latter, moreover,  decreases very fast like $\sim1/u^2$ when $u\to\infty$, it is  expected that the main
contribution results from the region around  $u\sim1.$  The  situation  is  somewhat different with respect to the variable $\rho$.  
While in this case  the integrand falls off for $\rho\to\pm \infty$ as well, it is a regular  function in $\rho$ with vanishing limit  at $\rho\to0.$
In order to elucidate the mainly contributing region of this integral,  we   split the integration domain  into three
regions: from $-\infty$ to $-\rho_0$ (lower  region),  from $-\rho_0$ to $\rho_0$ (inner  region) and from $\rho_0$
to $\infty$ (upper  region). This is, 
\[
\int_{-\infty}^\infty\frac{d\rho}{\rho}\ldots=\int_{-\infty}^{-\rho_0}\frac{d\rho}{\rho}\ldots+\int_{-\rho_0}^{\rho_0}\frac{d\rho}{\rho}\ldots+\int_{\rho_0}^\infty\frac{d\rho}{\rho}\ldots
\]
where the positive dimensionless parameter $\rho_0$ is chosen to fulfill the conditions
\begin{eqnarray}
\xi^{-1}\ll\rho_0\ll1\quad\mathrm{and}\quad
 (\lambda/\xi^2)^{\nicefrac{1}{3}}\ll\rho_0\ll1.\label{constraint}
\end{eqnarray}
Within the  inner narrow integration region, where $|\rho|\leqslant\rho_0\ll1$, one may Taylor expand $\eta$ in  the 
exponential and, separately, the remaining part of the integrand.  Afterwards,  we perform 
a change of  variable according to $s=\rho\xi$.  The latter is also carried out  in the integrals over the lower and upper  
regions but,  after extending the  resulting integration limit
$\rho_0\xi\to\infty,$ no relevant contribution comes from them. Thus, the total integral over $\rho$ can be well
approximated  by
\begin{eqnarray}\label{approax}
\int_{-\infty}^{\infty}\frac{d\rho}{\rho}\ldots\approx\int_{-\infty}^{\infty}ds\left\{2 s\sin\left[\frac{2u}{\lambda\xi}\left(s+\frac{s^3}{3}\right)\right]
+\frac{8u(u-1)}{3\lambda\xi}\right.\nonumber\\ \times\left.s^2\cos\left[\frac{2u}{\lambda\xi}\left(s+\frac{s^3}{3}\right)\right]\right\}.
\end{eqnarray} We remark that Eq.~(\ref{approax}) is accurate up to terms that decrease exponentially, like $\sim(\rho_0\xi)^{-1}\exp\left[-\frac{2u}{3\lambda\xi}(\rho_0\xi)^3\right]$ or even faster.

The integration over $s$  can, then,  be done  with the help of  the Macdonald function $K_{\nu}(x)$ using the relations\footnote{Equations (3.695.1-2)  in Ref. \cite{Gradshteyn} allow to find the  identity
\[
\int_{-\infty}^\infty dy \cos(by+ay^3)=\frac{2}{3} \sqrt{\frac{b}{a}}K_{\nicefrac{1}{3}}\left(\frac{2}{3\sqrt{3}}\frac{b^{\nicefrac{3}{2}}}{a^{\nicefrac{1}{2}}}\right).
\]After differentiating  with respect to $b$ and using Eqs. (8.486.10-11) of the aforementioned reference we  establish the relations (\ref{jeje1})-(\ref{jeje2}).}
\begin{eqnarray}\label{jeje1}
\int_{-\infty}^\infty dy\ y \sin(by +a y^3)=\frac{2}{3\sqrt{3}}\frac{b}{a}K_{\nicefrac{2}{3}} \left(\frac{2}{3\sqrt{3}}\frac{b^{\nicefrac{3}{2}}}{a^{\nicefrac{1}{2}}}\right),\\\label{jeje2}
\int_{-\infty}^\infty dy\ y^2 \cos(by +a y^3)=-\frac{2}{9}\frac{b^{\nicefrac{3}{2}}}{a^{\nicefrac{3}{2}}}K_{\nicefrac{1}{3}} \left(\frac{2}{3\sqrt{3}}\frac{b^{\nicefrac{3}{2}}}{a^{\nicefrac{1}{2}}}\right).
\end{eqnarray} We insert  the  resulting expression  into Eq. (\ref{modified}) and use  the   identity
$-x K_{\nicefrac{1}{3}}(x)=x\frac{d}{dx}K_{\nicefrac{2}{3}}(x)+\frac{2}{3}K_{\nicefrac{2}{3}}(x)$ (see Eqs. (8.486.12) and (8.486.16) in \cite{Gradshteyn}) to obtain
\begin{equation}\label{scalastrong}
\mathfrak{R}^{(0)}\backsimeq\frac{\alpha m^2}{12\sqrt{3} \pi\omega}\int_1^{\infty}\frac{du}{u\sqrt{u(u-1)}} K_{\nicefrac{2}{3}}\left(\frac{4u}{3\zeta}\right)(4u-1)
\end{equation}where, in addition, an integration by parts  has been carried out. Besides, the abbreviation  
$\zeta\equiv \lambda\xi=\frac{\omega}{2m}\frac{E}{E_c}(1-\hat{\pmb{k}}\cdot\hat{\pmb{\varkappa}})$ has been introduced, with the critical field strength 
$E_c=m^2/e$.  The parameter $\zeta$ encodes the two paths to vacuum polarization effects: either by increasing  the external field amplitude  or by 
increasing  the photon energy. It also depends on the propagation directions of the  photon $\hat{\pmb{k}}=\pmb{k}/\vert\pmb{k}\vert$ and 
the strong laser field $\hat{\pmb{\varkappa}}=\pmb{\varkappa}/\vert\pmb{\varkappa}\vert.$ Observe that  $\zeta$ is maximized  when the fields counterpropagate, 
i.e., when $\hat{\pmb{k}}\cdot\hat{\pmb{\varkappa}}=-1.$ 

We recall that Eq. (\ref{scalastrong}) was derived under  the assumptions given in (\ref{constraint}). Thus, the rate expression above   applies  when the 
number of absorbed laser photons is very large,  $n\geqslant n_*\approx\xi^2/\lambda\gg1.$  In this case, many photon orders $n$ contribute to the production 
rate \cite{Reiss1962,narozhnyi}. Note that Eq.~(\ref{scalastrong}) is  independent of the frequency of the strong background field.  The problem thus becomes 
quasi-static with respect to the external wave which, consequently, may be approximated by a pure constant crossed field.  We have checked that, starting from 
the   general  expression  of the vacuum polarization tensor  $\Pi_{\mu\nu}^{\left(\mathrm{scal}\right)}$ in such an external field configuration 
\cite{Schubert:2000yt}, we can also arrive at Eq.~(\ref{scalastrong}).

The corresponding production rate of  spin$-\frac{1}{2}$ particles emer\-ges  from  (\ref{scalastrong}) via the replacement:   $(4u-1)\to 2(8u+1).$ 
The resulting  integrand  turns  out to be  greater than the respective  one of Eq. (\ref{scalastrong}) within the whole integration domain of the 
$u-$variable. This fact  guarantees that $\mathfrak{R}^{\left(\frac{1}{2}\right)}/\mathfrak{R}^{(0)}>1$.  Hence the internal spin degrees of freedom 
promote the creation rate of spin-$\frac{1}{2}$ particles as compared with  scalar ones.

So far,  no restriction has been imposed on the parameter $\zeta.$ Let us consider the situation in which  $\zeta\gg1.$ To be  consistent  with our previous
condition ($\xi\gg1$) the parameter $\lambda$ must be restricted to $\lambda\gg 1/\xi$. 
We may  exploit the small-argument behavior of the Macdonald functions, $K_\nu(x)\sim\frac{\Gamma[\nu]}{2}\left(\frac{2}{x}\right)^\nu$ \cite{Gradshteyn}, and obtain
\begin{eqnarray}
\mathfrak{R}^{(0)}\backsimeq\frac{\alpha m^2 \zeta^{\nicefrac{2}{3}}}{6\sqrt{3\pi}\omega}\left(\frac{3}{2}\right)^{\nicefrac{2}{3}}\frac{\Gamma^2\left(\frac{2}{3}\right)}{\Gamma\left(\frac{13}{6}\right)},
\end{eqnarray}with $\Gamma(x)$ denoting the Gamma function. 
The corresponding result  for the fermionic case is given in  \cite{baierbook}.
%\begin{equation}
%\mathfrak{R}^{\left(\frac{1}{2}\right)}\backsimeq\frac{5\alpha m^2 \zeta^{\nicefrac{2}{3}}}{6\sqrt{3\pi}\omega}\left(\frac{3}{2}\right)^{\nicefrac{2}{3}}\frac{\Gamma^2\left(\frac{2}{3}\right)}{\Gamma\left(\frac{13}{6}\right)}.
%\end{equation} 
Here  the  role of the internal spin degrees of freedom are manifest as $\mathfrak{R}^{\left(\frac{1}{2}\right)}=5\mathfrak{R}^{(0)}.$

On the contrary, if $\zeta\ll1$ (corresponding to $\lambda\ll1$ at $\xi\gg1$), the large  asymptotic behavior of the  Macdonald functions
applies, i.e. $K_\nu(x)\sim \sqrt{\frac{\pi}{2 x}}e^{-x}$ \cite{Gradshteyn}.  With this expansion  in  mind, the  integration  over $u$ 
can be performed. The latter  becomes particularly simple because the region $u\sim1$ provides the essential contribution. This leads to
\begin{equation}
\mathfrak{R}^{(0)}\backsimeq\frac{\alpha m^2 \zeta}{16\omega}\sqrt{\frac{3}{2}}e^{-\frac{4}{3\zeta}}.
\end{equation}A comparison with the corresponding fermionic rate \cite{Greiner}  leads to write  $\mathfrak{R}^{\left(\frac{1}{2}\right)}=6\mathfrak{R}^{(0)}.$  It is interesting to note 
that a similar result was found for the strong-field Bethe-Heitler process  $\gamma_{\rm{Coul}}+n\gamma_L\to \ell^-+\ell^+.$ In the 
parameter domain where $\xi\gg1$ and $E\sim E_c$, the rate associated with the creation of Dirac fermions in the Coulomb field of a 
nucleus  exceeds by  a factor of $\simeq7$  the corresponding rate for spin$-0$ particles 
\cite{Muller:2010xq}.

%%%%%%%%%%%%%%%%%%%%%%%%%%%%%%%%%%%%%%%%%%%%%%%%%%%%%%%%%%%%%%%%%%%%%%%%%%%%%%%%%%%%%%%%%%%%%%%%%%%%%%
\section{Conclusions and outlook}
%%%%%%%%%%%%%%%%%%%%%%%%%%%%%%%%%%%%%%%%%%%%%%%%%%%%%%%%%%%%%%%%%%%%%%%%%%%%%%%%%%%%%%%%%%%%%%%%%%%%%%

Photo-initiated  production of a pair of spinless particles in the  field 
of a circularly polarized laser wave was investigated. Compact expressions for the
pair production rate were obtained in various asymptotic parameter regimes.
Our analysis  was carried out by considering the imaginary part of the  
vacuum polarization tensor as dictated  by the optical theorem.
Comparisons between the creation rates for spin-$\frac{1}{2}$ versus 
spin-$0$ particles revealed how spin effects are manifest at different 
energy scales. It was shown that the spin degrees of freedom affect the 
absolute magnitude and, in the limit of high photon energies, also the functional
form of the production rates. The rate predictions based on the Dirac theory were 
generally found to be significantly larger than the corresponding results for 
Klein-Gordon particles.

We note that the absence of internal spin degrees of freedom  in the scalar theory 
renders the evaluation of the  vacuum polarization  tensor somewhat easier 
than in the fermionic case. It is worth mentioning that, since the total production rates 
for spin-$0$ and spin-$\frac{1}{2}$ particles in the high-field limit ($\xi\gg 1$) 
differ by an overall factor only, this technical simplification could be exploited to extract 
new insights into the production process which are valid not only for scalar particles 
but also for $e^-e^+$ pairs. Indeed, the very similar behavior of the production rates
in this regime indicates that the process is mainly determined by
the features of the weak photon and strong laser fields, rather than  the particular 
property of the  matter field.

Finally we point out that,  whenever the energy scale remains
within the phenomenological  limits of  QED and  its
fundamental principles  are preserved, the  
photo-production of a hypothetical pair of spinless
particles  characterized by  a tiny fraction $\epsilon$ of the
electron charge $e$ would  not differ qualitatively
from the  respective creation of a  spinless electron-positron pair. In
correspondence, the rate associated with  the latter
phenomenon can be obtained from the  expression derived  in this
Letter by  replacing the  electron  parameters
$(e,m)$ by the respective quantities $(\epsilon e, m_\epsilon)$
associated with a minicharged particle. 
Therefore, our results can be useful in the ongoing search
for minicharged  particles \cite{Gies:2008wv,Jaeckel:2010ni,Dobrich:2012sw,Ahlers:2006iz} 
and help to improve our understanding of how the spin degrees of freedom 
affect the relevant experimental observables.

\vspace{0.005 in}
\begin{flushleft}
\textbf{Acknowledgments}
\end{flushleft}
\vspace{0.005 in}
S. Villalba-Ch\'avez gratefully acknowledges the support by the Alexander von Humboldt Foundation.

\end{document}